\documentclass[aps,prl,twocolumn,superscriptaddress,showpacs,floatfix,nobibnotes,pdftex]{revtex4}

\usepackage{subfigure}
\usepackage{epsfig}
\usepackage{epstopdf}
\usepackage{color}
\usepackage{graphicx}
\usepackage{longtable}
\usepackage{CJK}
\usepackage{amsmath}
\usepackage{float}
\newcommand{\blue}[1]{{ \color{blue} #1 }}
\ifx\pdfoutput\undefined
\usepackage[dvips]{graphicx}
\usepackage[dvipdfm,
  pdfstartview=FitH,
     CJKbookmarks=true,
          bookmarksnumbered=true,
          bookmarksopen=true,
         colorlinks=true,
          pdfborder=002,
          citecolor=blue%
           ]{hyperref}
\AtBeginDvi{} 
\else
\usepackage[
          pdfstartview=FitH,
          CJKbookmarks=true,
          bookmarksnumbered=true,
           bookmarksopen=ture,
           colorlinks=true,
           citecolor=blue
           ]{hyperref}

\begin{document}

\title{A nucleon-pair and boson coexistent description of nuclei}

\author{Lianrong Dai}
\affiliation{Department of Physics, Liaoning Normal University,
Dalian 116029, China}

\author{Feng Pan\footnote{The corresponding author's e-mail: \blue{daipan@dlut.edu.cn}}}
\affiliation{Department of Physics, Liaoning Normal University,
Dalian 116029, China}\affiliation{Department of Physics and
Astronomy, Louisiana State University, Baton Rouge, LA 70803-4001,
USA}

\author{J. P. Draayer}
\affiliation{Department of Physics and Astronomy, Louisiana State
University, Baton Rouge, LA 70803-4001, USA}

\begin{abstract}
We study a mixture of $s$-bosons and like-nucleon pairs
with the standard pairing interaction outside an inert core.
Competition between the nucleon-pairs and  $s$-bosons
is investigated in this scenario.
The robustness of the BCS-BEC coexistence and crossover phenomena are examined through an analysis
of $pf$-shell nuclei with realistic single-particle energies, in which two configurations with Pauli
blocking of nucleon-pair orbits due to the formation of the $s$-bosons is taken into account.
When the nucleon-pair orbits are considered to be independent of the $s$-bosons,
the BCS-BEC crossover becomes smooth, with the number of the
$s$-bosons noticeably more than that of the nucleon-pairs near the half-shell point,
a feature that is demonstrated in the $pf$-shell
for several values of the standard pairing interaction strength.
{As a further test of the robustness of the BCS-BEC coexistence
and crossover phenomena in nuclei, results are given for}
B(E2; $0^{+}_{1}\rightarrow 2^{+}_{1}$) values of even-even $^{102-130}$Sn
{with $^{100}$Sn taken as a core} and valence neutron pairs confined
within the $1d_{5/2}$, $0g_{7/2}$,
$1d_{3/2}$, $2s_{1/2}$, $1h_{11/2}$
orbits in the nucleon-pair orbit and the $s$-boson independent approximation.
The results indicate that the B(E2) values are reproduced well.
\end{abstract}

\maketitle
\noindent{\bf Introduction:}
{The interacting boson model (IBM)~\cite{1}, like the collective model~\cite{2} of Bohr and Mottelson and
the microscopic shell model~\cite{31,3}, has been successful in providing a logical framework for studying the
structure of atomic nuclei, at least in the low-energy regime. Many studies show that the $J=2$ $d$-bosons of
the IBM are similar to the $d$-phonons that emerge naturally from a five-dimensional harmonic oscillator
description of the quadrupole vibrations of the collective model. The addition of a $J=0$ $s$-boson to the
$d$-boson picture allows the IBM to accommodate couplings to low-energy monopole modes, which expands the
U(5) algebra structure that underpins the five-dimensional oscillator to a six-dimensional U(6) algebra~\cite{4}.
Within this expanded U(6) framework of the IBM, the total number of bosons is regarded as the number of
valence particle or hole pairs \cite{5,6}.}
{It is now commonly accepted that there is a close relationship between the $s$-bosons and
$J=0$ pairs of like-nucleons, since pairs of fermions in these systems often exhibit boson-like behavior \cite{7,8}.
It has been shown  that bosons may emerge from fermionic pairing
due to spontaneous symmetry breaking of the Bardeen-Cooper-Schrieffer (BCS) type~\cite{81,82}.
However, a schematic study of the relative roles played by $J=0$ nucleon pairs and
$s$-bosons as employed in the IBM in nuclei is still lacking.}

\vskip .3cm
Generally, the pairing correlation in Fermi many-body systems
can be understood in terms of attractive interactions among fermion pairs
manifested by the BCS mechanism.
When the attractive interaction between two fermions is strong enough,
on the other hand,
the two fermions may form a bosonic bound state with Bose-Einstein
condensation (BEC) in the ground state of the system.
As shown in ~\cite{9,10,11,12,13,duk},
a smooth crossover from a Cooper-paired state to a Bose condensate state
of tightly bound pairs takes place in the continuum model of a Fermi gas at zero temperature.
In \cite{14}, the coexistence of the BCS and the BEC-like
pair structures in   $^{11}$Li was investigated based on phase space analysis.
Inspired by the {above developments,}
we consider a mean-field plus standard pairing model including
$s$-bosons, where the $s$-bosons are regarded as tightly bound nucleon pairs,
which should also emerge naturally in nuclei according to the aforementioned observations.

\vskip .3cm
\noindent{\bf The model:}
Since some nucleon-pairs in a nucleus may behave like bosons,
we consider a mixture of $s$-bosons and like-nucleon pairs
with the standard pairing interaction outside an inert core.
A schematic Hamiltonian considered is given by
\begin{eqnarray}\label{H}\nonumber
&\hat{H} =\sum_{j}\epsilon_{j}\hat{N}_{j}-G\sum_{j,\,j'}S^{+}_{j}S^{-}_{j'}+\\
&\alpha(\hat{n}_{s})-r\,G\,\sqrt{\Omega}\sum_{j}(S^{+}_{j}s+s^{\dagger} S^{-}_{j}),
\end{eqnarray}
where $j$ and $j'$ run over $p$ distinct orbits considered,
$s^{\dagger}$ ($s$) is the $s$-boson creation (annihilation) operator,
$\{\epsilon_{j}\}$  is a set of single-particle energies
generated from any mean-field theory,
$\hat{N}_{j}=\sum_{m}a^{\dagger}_{j m}a_{j m}$
and $S_{j}^{+}=\sum_{m>0}(-1)^{j-m}a^{\dagger}_{jm}a^{\dagger}_{j-m}$
($S_{j}^{-}=(S_{j}^{+})^{\dagger}$),
in which $a^{\dagger}_{jm}$ ($a_{jm}$)
is the creation (annihilation) operator
for a nucleon with angular momentum quantum number $j$ and that of its projection $m$,
$G>0$ is
the overall pairing interaction
strength, $r$ is a scale factor used to
describe the interaction between $s$-boson and nucleon pairs,
$\Omega=\sum_{j}(j+1/2)$ is the maximum pair-occupancy  of the $p$-orbit system,
and the pure $s$-boson part $\alpha(\hat{n}_{s})$, which is a function of the $s$-boson number operator
$\hat{n}_{s}=s^{\dagger}s$,
is adjusted to reproduce the ground state energy of the Hamiltonian with the first two terms
of the nucleon-pair sector for given $G$.
With this assumption, the ground state energy of the system in either
the pure BCS phase or the pure BEC phase is exactly the same.
In (\ref{H}), the interaction between the $s$-boson and the nucleon pairs is the same
as that among nucleon pairs when $r=1$.
The factor $\sqrt{\Omega}$ appears in the last term of (\ref{H}) because
the nucleon pair operator  $S^{+}\sim\sqrt{\Omega}s^{\dagger}$,
where $S^{+}=\sum_{j}S^{+}_{j}$ and $S^{-}=(S^{+})^{\dagger}$,
when the number of nucleon-pairs $k$ is far smaller than  $\Omega$
with $k\ll\Omega$.
Here and in the following, we assume that all the nucleons in the system are paired and that there is an even number of nucleons. It is obvious that the total number of the
$s$-bosons and that of the
nucleon pairs is a conserved quantity in the model.

\vskip .3cm
Since the $s$-bosons originate from nucleon-pairs,
many nucleon-pair orbits should be blocked due to the
{Pauli Exclusion Principle. In a single-$j$ case for example,
when there are $n_{s}$ $s$-bosons, the maximal nucleon-pair occupancy
should become $\tilde{\Omega}_{j}=j+1/2-n_{s}$.
Thus, for given $n=k+n_{s}\leq \Omega$,
the eigenstates of (\ref{H}) may be written as
\begin{equation}\label{2}
\vert n;\zeta\rangle=\sum_{k=0}^{n}\sum_{\xi_{k}}C^{(\zeta)}_{k,\xi_{k}}\vert n-k; \xi_{k},
k\rangle \vert n-k\rangle_{s},
 \end{equation}
where $\zeta$ labels the $\zeta$-th excitation state of the system,
$\vert n-k; \xi_{k}, k\rangle$ is the eigenstate of the mean-field plus standard
pairing model of the nucleon-pair sector with $\tilde{\Omega}=\Omega-(n-k)$.

\vskip .3cm
Generally, if there are $p$ orbits, {the maximum number of distinct blocking patterns is}
equal to the number of partitions of $n-k$ into $p$ integers $[\mu_{1},\cdots,\mu_{p}]$
with $\sum_{i=1}^{p}\mu_{i}=n-k$ and $0\leq\mu_{i}\leq\Omega_{j_{i}}$ for $i=1,\cdots,p$,
which results in the effective maximal nucleon-pair occupancy of each orbit
for a given partition $[\mu_{1},\cdots,\mu_{p}]$
to be $\tilde{\Omega}_{j_{i}}=\Omega_{j_{i}}-\mu_{i}$ for $i=1,\cdots,p$.
We take the $p=2$ case with $j_{1}=3/2$ and $j_{2}=5/2$ as an example.
If there is no $s$-boson with $n_{s}=0$, the maximal pair occupancy
of each orbit is given by $\Omega_{1}=2$ and $\Omega_{2}=3$, respectively.
When $n_{s}=1$, there are two blocking configurations. The first has
$\tilde{\Omega}_{1}=1$ and $\tilde{\Omega}_{2}=\Omega_{2}=3$,
for which the $s$-boson is formed from two nucleons in the first orbit. The second has $\tilde{\Omega}_{1}={\Omega}_{1}=2$ and $\tilde{\Omega}_{2}=2$,
for which the $s$-boson is formed from two nucleons in the second orbit.
Let the nucleon-pair product states be denoted as $\vert n_{1};n_{2}\rangle$,
where $n_{1}$ and $n_{2}$ are the number of pairs
with $0\leq n_{1}\leq 2$ and $0\leq n_{2}\leq 3$, which should all be needed
in diagonalizing the first two terms of (\ref{H}) when $n_{s}=0$.
However,  $\vert n_{1}=2;n_{2}=0\rangle$ should be ruled out
in diagonalizing the first two terms of (\ref{H}) for $n_{s}=1$
in the first blocking configuration, while
$\vert n_{1}=0;n_{2}=3\rangle$ should be ruled out
in diagonalizing the first two terms of (\ref{H}) for $n_{s}=3$
in the second blocking configuration. Namely, the dimension
of the subspace spanned by the nucleon-pair product states
will be reduced due to Pauli blocking.

\vskip .3cm
The eigenstates of the first two terms of (\ref{H}),
$\vert n-k; \xi_{k}, k\rangle$,
appear in (\ref{2}) with a given number of nucleon-pairs $k$ for
distinct blocking patterns. However, they are not linearly independent since they are all $k$-pair states.
In order to treat this situation more precisely, we need to
introduce a mixed state description if the probability
of a given partition of the blocking is known.
In this work, we still treat {this feature within a pure} state description for simplicity.
Hence, only one of the blocked configurations for a given $k$ is considered.

\vskip .3cm
{As a relatively simple example, we set $r=1$ and consider a $pf$-shell system with
the $4$ orbitals $0f_{7/2}$, $1p_{3/2}$, $1p_{1/2}$,  and $0f_{5/2}$ assigned the
single-particle energies deduced in
\cite{HOB}; that is, we set $\epsilon_{7/2} =-8.624$\,MeV,
$\epsilon_{3/2} =-5.6793$\,MeV, $\epsilon_{1/2} =-4.137$\,MeV,
and $\epsilon_{5/2} =-1.3829$\,MeV in the Hamiltonian (\ref{H}).
Then the eigen-equation used to determine the
eigen-energies $E^{(\zeta)}_{n}$ of (\ref{H}) and the corresponding expansion coefficients $C^{(\zeta)}_{k,\xi_{k}}$
of (\ref{2}) are given by}
\begin{eqnarray}\label{3}\nonumber
&{1\over{G}}\left(E_{k}^{(\xi_{k})}+\alpha(n-k)- {E}^{(\zeta)}_{n}\right)
C^{(\zeta)}_{k,\xi_{k}}=\\\nonumber
&\sqrt{\Omega(n-k+1)}\sum_{\xi_{k-1}}C^{(\zeta)}_{k-1,\xi_{k-1}}\times\\\nonumber
&{\small\langle n-k; \xi_{k}, k\vert S^{+}\vert n-k+1; \xi_{k-1}, k-1\rangle}\\\nonumber
&+\sqrt{\Omega(n-k)}\sum_{\xi_{k+1}}C^{(\zeta)}_{k+1,\xi_{k+1}}\times\\
&\langle n-k; \xi_{k}, k\vert S^{-}\vert n-k-1; \xi_{k+1}, k+1\rangle
\end{eqnarray}
for $k=0,~1,~\cdots,~n$,
where  $E^{(\xi_{k})}_{k}$ is the $\xi_{k}$-th $k$-pair
excitation energy of the original mean-field plus standard pairing Hamiltonian
of the nucleon-pair sector with $\tilde{\Omega}=\Omega-n+k$ for
a fixed  partition $[\mu_{1},\cdots,\mu_{p}]$ of the blocking,
{and} the matrix elements
$\langle n-k;\xi_{k}, k\vert S^{+}\vert n-k+1;\xi_{k-1}, k-1\rangle=
\langle n-k+1; \xi_{k-1}, k-1\vert S^{-}\vert n-k;\xi_{k}, k\rangle^{*}$
can be calculated when all excited states of the original
mean-field plus standard pairing model for the nucleon-pair sector
are obtained.
For a given number of nucleon-pairs $k$,
we calculate all excitation energies and the corresponding
eigenstates
of the original mean-field plus standard pairing Hamiltonian
of the nucleon-pair sector with $\tilde{\Omega}=\Omega-n+k$
for a fixed partition of the blocking, which can be done by using the
the Heine-Stieltjes polynomial approach~\cite{guan}.

\begin{figure}[htp]
\begin{center}
\includegraphics[scale=0.142]{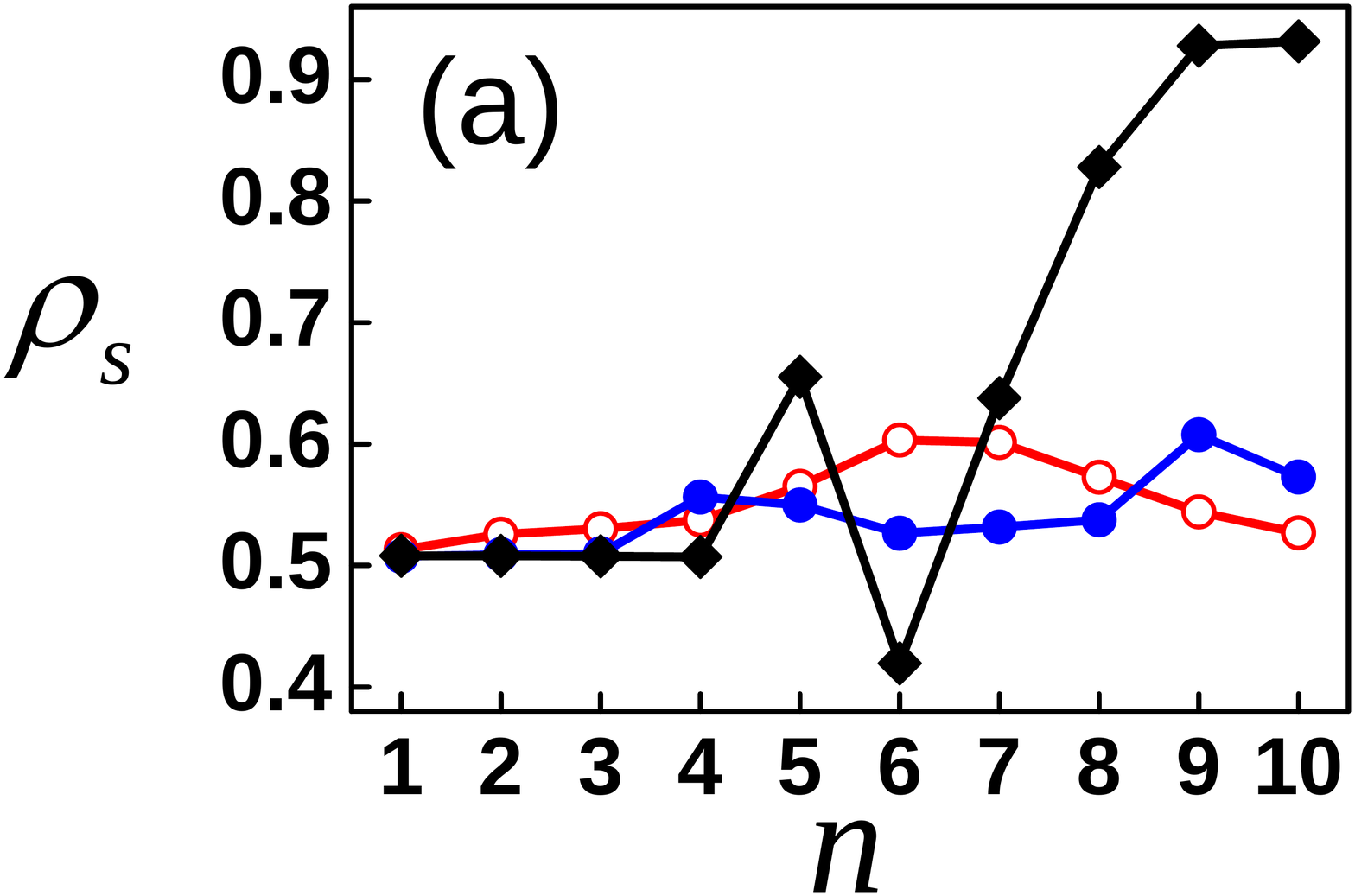}
\includegraphics[scale=0.142]{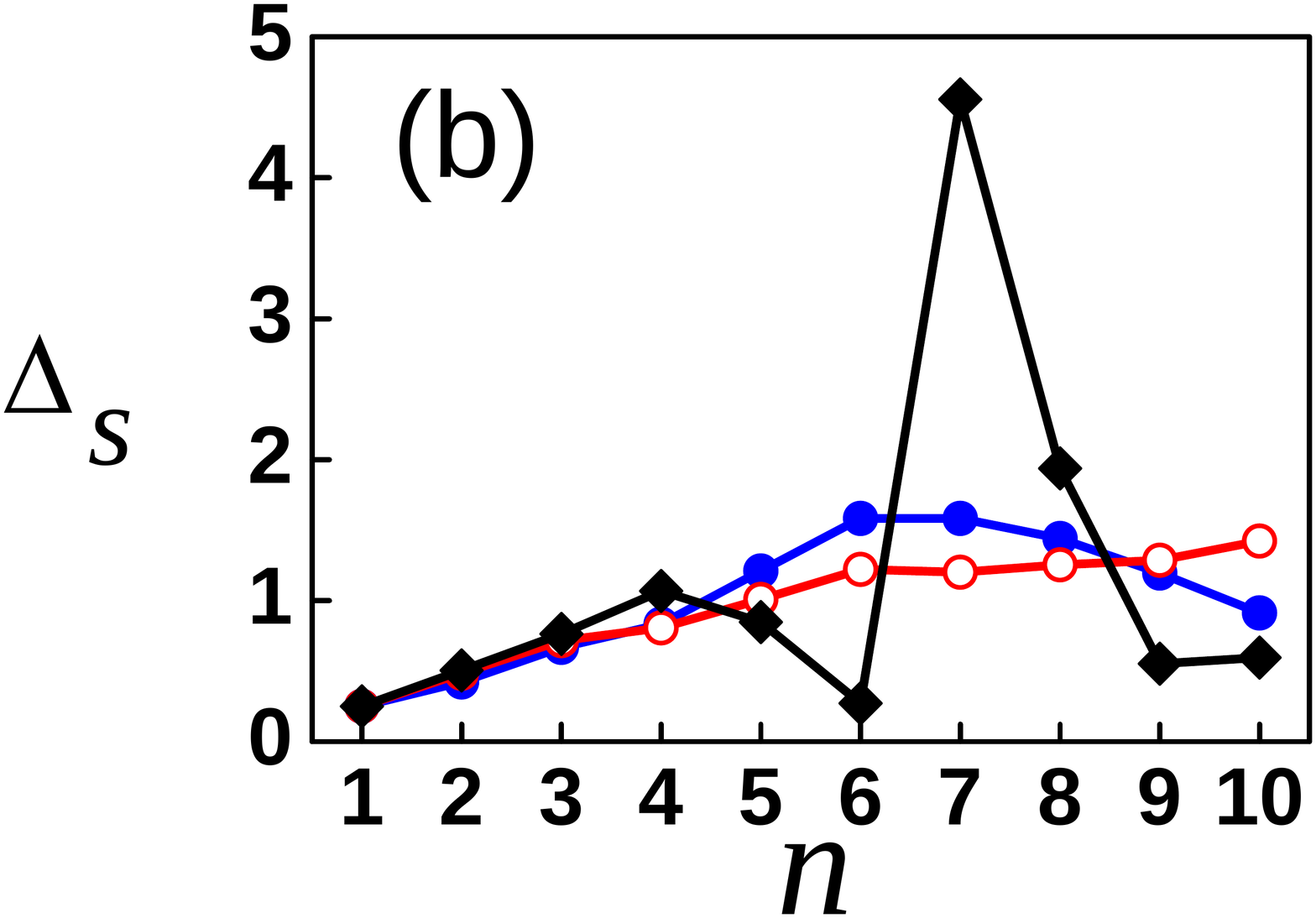}
\caption{(Color online) The ground-state $s$-boson occupation probability $\rho_{s}$ {(a)}
and the ground-state $s$-boson number fluctuation  $\Delta_{s}$ {(b)}
as functions of $n$ for different blocking configurations considered
in the $pf$-shell with $r=1$ and $G=1.4$\,MeV.
The solid diamonds denote the results of the first blocking configuration,
the solid dots denote the results of the second configuration,
and the open circles denote the results with the nucleon-pair orbit and the $s$-boson
independent approximation.
\label{fig1}}
\end{center}
\end{figure}

\vskip .3cm

{After} all excitation energies $\{E^{(\xi_{k})}_{k}\}$
and the corresponding eigenstates of the nucleon-pair part of the model
are obtained for $1\leq k\leq\Omega$ with $\Omega=10$ for this case,
{we simply set} $\alpha(n_{s})=E^{(\xi_{k}=1)}_{k}$ with $k=n_{s}$
for a given value of $G$.
The orbits are {then} arranged in order according to
the values of the single-particle energies with $\epsilon_{j_{1}}<\cdots<
\epsilon_{j_{p}}$.
Specifically, we consider two blocking configurations.
In the first,
the blocking to the nucleon-pair orbits results in the configuration with the
lowest orbits blocked,
namely, we take $\mu_{1}$ to be the largest possible integer,
then $\mu_{2}$ to be the remaining largest possible integer, and so on.
In the second configuration, the blocking to the nucleon-pair orbits results in the configuration with the
highest orbits {blocked; namely,} we take $\mu_{p}$ to be the largest possible integer, then $\mu_{p-1}$ to be the remaining largest possible integer, and so on.
More precisely, in the first blocking configuration,
when $n_{s}\leq 4$, we assume the $j_{1}=7/2$ orbit is blocked
because it is the lowest in energy. When $5\leq n_{s}\leq 6$,
we assume the $j_{1}=7/2$  and $j_{2}=3/2$ orbits are blocked, and so on.
Similarly, in the second blocking configuration, when $n_{s}\leq 3$,
we assume the $j_{4}=5/2$ orbit is blocked because it is the highest in energy.
When $n_{s}=4$, we assume the $j_{4}=5/2$  and $j_{3}=1/2$ orbits
are blocked, and so on.
\vskip .3cm
In the analysis, the pairing interaction strength $G=1.4$\,MeV is fixed.
Once (\ref{3}) is {solved numerically,} we calculate  the ground-state
occupation probability $\rho_{s}$ of the $s$-bosons
\begin{equation}\label{5}
\rho_{s}={1\over{n}}\langle n;\zeta=1\vert\hat{n}_{s}\vert n;\zeta=1\rangle
 \end{equation}
and the ground-state $s$-boson number fluctuation defined by
\begin{equation}\label{6}
\Delta_{s}=\sqrt{\langle n;\zeta=1\vert\left(\hat{n}_{s}-n\rho_{s}\right)^{2}\vert n;\zeta=1\rangle}.
\end{equation}

As shown in  Fig.~\ref{fig1}(a), the $s$-boson
occupation probability $\rho_{s}$ is
always greater than $40\%$, increasing from $40\%$ to $93\%$
for $n=6$ to $n=10$ in the first blocking configuration. In the second blocking configuration it is always greater than $50\%$, increasing from $50\%$ to $60\%$ over the range of $n$ values shown.
The staggering in the occupation probability
happens near the half-filling point in the first blocking configuration. In both
cases, the $s$-boson occupation probability reaches its maximal value
near the shell closure point.

\vskip .3cm
The results for the nucleon-pair orbit and the $s$-boson
independent approximation are also shown in {Fig. \ref{fig1}} for comparison,
with the eigenstates of the mean-field plus standard
pairing model of the nucleon-pair sector $\{\vert n-k; \xi_{k}, k\rangle\}$
used in (\ref{2}) replaced by  $\{\vert 0; \xi_{k}, k\rangle\}$
for any $n$ and $k$ with
the maximal nucleon-pair occupancy of each orbit unchanged,
namely, the blocking to the nucleon-pair orbits due to the formation
of the $s$-bosons is not considered.
In this case, there is a critical region of the BCS-BEC crossover, which  tracks with
a range of $n$ values near the half-filling of the shell where
$\rho_{s}$ reaches its  maximal value.
Though the BCS-BEC crossover behavior in these three cases are different,
the BCS-BEC coexistence seems robust.
{In any case,} once the $s$-bosons
emerge in the system with the strong pairing interaction
shown in~\cite{11,12,13,duk}, which prefers to occur in a dilute fermion-pair
environment~\cite{11,12,13,duk,14}, the $s$-boson content, in general,
changes noticeably  with increasing $n$
due to the interaction between nucleon-pairs and the $s$-bosons.
This reinforces the BCS-BEC crossover even when the number of nucleon-pairs becomes
large.
In addition, as further shown in Fig. \ref{fig1}(b), the $s$-boson number fluctuation
also changes rapidly with $n$, indicating that the nucleon-pair constituent is also
significant, which becomes more noticeable near the half-filling point.
Therefore, once the $s$-boson emerges in the system,
the pure BEC phase never occurs except in the first blocking configuration in the large $\Omega$-limit.
Rather, the system always seems to be in a
BCS-BEC coexistence phase,  which is robust for any finite $\Omega$ and $n$,
a feature that is consistent with
the conclusion made in \cite{14}, with the $s$-boson content
greater than that of the nucleon-pairs in general.
Thus, the $s$-boson formation induced by the pairing interaction with fewer nucleon-pairs
and the further BCS-BEC crossover enhancement with increasing
nucleon-pairs seems to track with the emergence of $s$-bosons in a nucleus.

\vskip .3cm
{Since we currently lack probability distribution information
for} different blocking configurations, the nucleon-pair orbit and the $s$-boson
independent approximation is adopted in the following.
The $s$-boson occupation probability $\rho_{s}$ and the
$s$-boson number fluctuation in the $pf$-shell for
$G=0.2$\,MeV, $0.6$\,MeV, $1.0$\,MeV, and $2.0$\,MeV
with the approximation are shown in Fig. \ref{fig2},
in which the scale factor $r=1$ is still taken.
It should be noted that
$G$ ranges from $0.4\sim0.7$\,MeV in the $pf$-shell
in order to reproduce the GXPF1 $J=0$ and $T=1$
pairing excitation spectrum~\cite{pan17}.
Hence, the results shown in Fig. \ref{fig2}
not only  provide information about realistic
situations, but also show results for  a slightly
weaker pairing interaction  with $G=0.2$\,MeV and
a stronger pairing interaction with $G=2.0$\,MeV.
As shown in Fig. \ref{fig2}(a), similar to the
case shown in Fig. \ref{fig1},
the ground-state $s$-boson occupation probability
$\rho_{s}$ is always greater than $50\%$.
There is a critical region of the BCS-BEC crossover, which
tracks with a range of $n$ values near the half-filling of the shell
where $\rho_{s}$ reaches its  maximal value.
The critical region moves towards a
larger $n$ interval with $n>\Omega/2$ when the pairing interaction
becomes stronger. It is also clear that
the maximal value of $\rho_s$ increases with increasing
$G$, and is about $62\%$ when $G=2.0$~MeV.
{As can be seen} in Fig. \ref{fig2}(b),
the largest fluctuation $\Delta_{s}$ also occurs
in the critical region, within which
$\rho_{s}$ reaches its maximal value. Actually,
both the $s$-bosons and
the nucleon-pairs are most non-localized
in the crossover region in this case.

\begin{figure}[htp]
\begin{center}
\includegraphics[scale=0.1439]{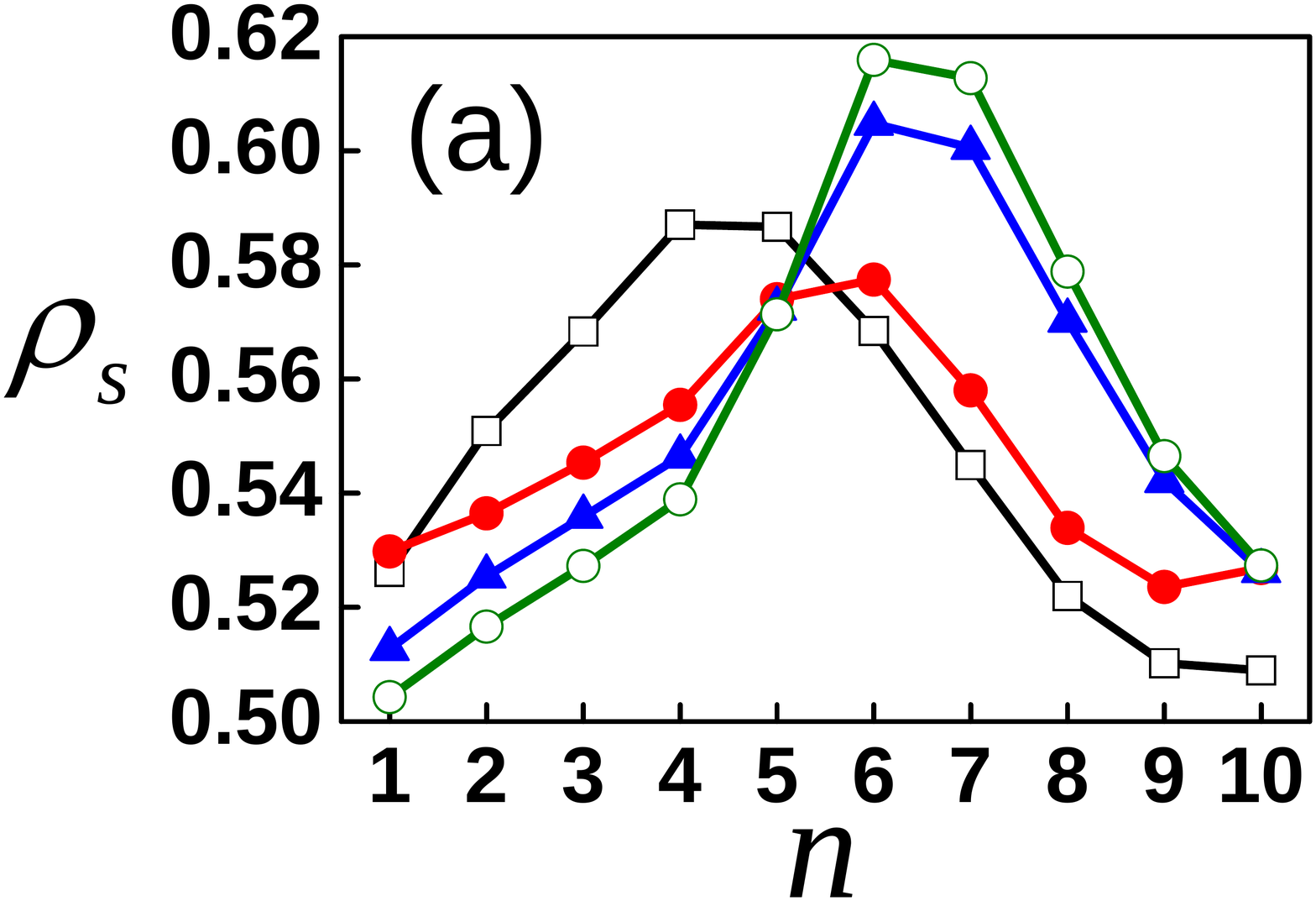}
\includegraphics[scale=0.1439]{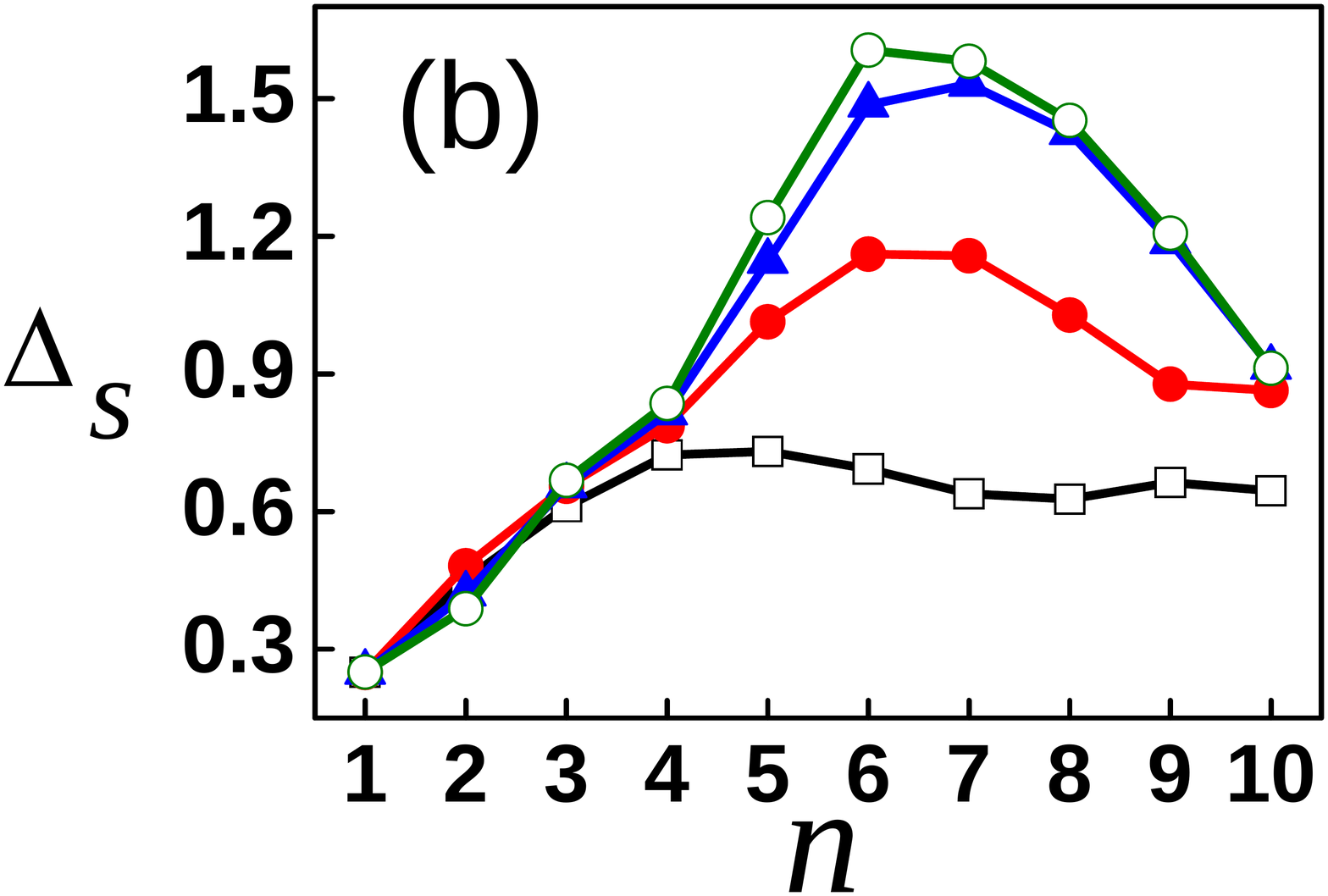}
\caption{(Color online) The same as Fig. \ref{fig1}, but for
different $G$ values in the nucleon-pair orbit and the $s$-boson
independent approximation.
The open squares denote the results with $G=0.2$\,MeV,
the solid dots denote the results with $G=0.6$\,MeV,
the solid triangles denote the results with $G=1.0$\,MeV,
and the open circles denote the results with $G=2.0$\,MeV.
\label{fig2}}
\end{center}
\end{figure}


\noindent{\bf An example of application:}
As an example of an application
of the theory with the nucleon-pair orbit and the $s$-boson
independent approximation,
we use the model to fit $2^{+}_{1}$ level energies and
then to calculate B(E2; $0^{+}_{1}\rightarrow 2^{+}_{1}$) values
of even-even  $^{102-130}$Sn, which have
attracted a lot of attention both
experimentally and theoretically \cite{be20,be21,be22,be23, th1, th2}.
In our calculation, $^{100}$Sn is
considered as the core of these nuclei with valence
neutron pairs confined
to the $1d_{5/2}$, $0g_{7/2}$,
$1d_{3/2}$, $2s_{1/2}$, and $1h_{11/2}$ 
orbits with single-particle energies $\epsilon_{5/2} =0.00$\,MeV,
$\epsilon_{7/2} =0.08$\,MeV, $\epsilon_{3/2} =1.66$\,MeV,
$\epsilon_{1/2} =1.55$\,MeV, and $\epsilon_{11/2} =3.55$\,MeV,
respectively~\cite{be20}. Since the low-lying states in the Sn isotopes
are almost spherical, the bosonic part of the Hamiltonian of (\ref{H})
in this case is chosen as the IBM U(5) type with $\alpha(\hat{n}_{s},\hat{n}_{d})=
\alpha(\hat{n}_{s})+\beta(\hat{n}_{d})$,
where $\beta(\hat{n}_{d})$ is a function of the number of $d$-bosons $\hat{n}_{d}$,
and $\alpha(n_{s})=E^{(\xi_{n_{s}}=1)}_{n_{s}}$ is still assumed.
The total number of bosons and nucleon pairs, $n=n_{s}+n_{d}+k$,
is a conserved quantity in this case.
For a given $n$,
the ground state is given by $\vert n,0^{+}_{1}\rangle=\vert n,\zeta=1\rangle$
determined by (\ref{2}), while the $2^{+}_{1}$ state, for simplicity,
is assumed to be one $d$-boson state, of which
the $d$-boson is assumed to be formed from two neutrons in the lowest
$1d_{5/2}$ orbit. Thus, the $2^{+}_{1}$ state
$\vert n, 2^{+}_{1},M\rangle=d^{+}_{M}\vert n-1,\zeta=1\rangle$,
where the Pauli blocking in the $1d_{5/2}$
orbit is considered in $\vert n-1,\zeta=1\rangle$ for $\vert n, 2^{+}_{1},M\rangle$.
{Furthermore,} the $2^{+}_{1}$ level energy is simply
given by $E(n,\, 2^{+}_{1})=\beta(1)-\beta(0)+{E}^{(1)}_{n-1}-E^{(1)}_{n}$ for a given $n$,
where ${E}^{(1)}_{n-1}$ is determined by (\ref{3}) with the Pauli blocking
in the $1d_{5/2}$ orbit considered.
In the fitting, the parameters $G$ and $\beta(1)-\beta(0)$ in the model
are fixed for all nuclei considered
with $G=0.18$\,MeV and $\beta(1)-\beta(0)=-10.617$\,MeV, while the scale factor
$r$ is adjusted to reproduce the experimental
$2^{+}_{1}$ level energy exactly, of which the values for these nuclei
are given in Table \ref{tb1}. It is clearly shown that $r$ deceases
almost linearly as the total number of bosons and nucleon pairs, $n$, increases.

\begin{table}[htp]
\caption{
The parameter $r$ used in fitting  the $2^{+}_{1}$ level energies (in MeV)
of even-even $^{102-130}$Sn, for which the experimental values shown in \cite{exp2} are used.}
\begin{center}
\begin{tabular}{ccccccccccccccccccccccccccccc}
 \hline
\toprule Nucleus &$n$ & {$r$} &&{$E(2^{+}_{1})$}&\vline& Nucleus & $n$& {$r$} &&{$E(2^{+}_{1})$}\\
\hline
$^{102}$Sn &1 & 3.885 &&1.472 &\vline  &$^{104}$Sn &2&3.544 && 1.260 \\
$^{106}$Sn &3& 3.270 &&1.208 &\vline &$^{108}$Sn  &4&3.025 && 1.206 \\
$^{110}$Sn &5& 2.792 &&1.212 &\vline  &$^{112}$Sn &6&2.579 &&1.257 \\
$^{114}$Sn &7& 2.366 &&1.300 &\vline  &$^{116}$Sn &8&2.127 &&1.294 \\
$^{118}$Sn &9& 1.862 &&1.230 &\vline  &$^{120}$Sn &10&1.585 &&1.171 \\
$^{122}$Sn &11& 1.318 &&1.141 &\vline  &$^{124}$Sn &12&1.063 &&1.132 \\
$^{126}$Sn &13& 0.822 &&1.141 &\vline  &$^{128}$Sn &14 &0.591 &&1.169 \\
$^{130}$Sn &15& 0.367 &&1.221 &\vline\\
\hline
\end{tabular}
\end{center}
\label{tb1}
\end{table}
The effective E2 operator in this case is defined as
\begin{eqnarray}\label{9}\nonumber
T_{\mu}({\rm E2})=q_{2}\left({1-\chi\over{2}}
(d^{\dagger}_{\mu}s+s^{\dagger}\tilde{d}_{\mu})\right.\\
\left.
+{1+\chi\over{2\sqrt{\Omega}}}(d^{\dagger}_{\mu}S^{-}+S^{+}\tilde{d}_{\mu})\right),
 \end{eqnarray}
where $d_{\mu}^{\dagger}$ is the $d$-boson creation operator,
$\tilde{d}_{\mu}=(-1)^{\mu}d_{-\mu}$, in which
$d_{\mu}$ is the $d$-boson annihilation operator,
$q_{2}$ is the effective quadrupole parameter,
and $\chi\in[-1,1]$ is used. In the fitting,
$\chi=0.44$ and $q_{2}=0.0827$\,eb are chosen
from the best fit for all nuclei considered.
The  B(E2; $0^{+}_{1}\rightarrow 2^{+}_{1}$) obtained from this
theory and the corresponding experimental values, together
with the results obtained from the large-scale shell
model (LSSM) with the same shell model space consideration \cite{be20},
are shown in Fig. \ref{fig3}(a), which indicates that the experimental data
are well reproduced by this theory with the parameter $r$ determined
uniquely by the $2^{+}_{1}$ level energy, except that the B(E2) values of $^{116}$Sn
and $^{130}$Sn are a little larger than the corresponding experimental
values. Overall, the B(E2; $0^{+}_{1}\rightarrow 2^{+}_{1}$)
values for even-even $^{102-114}$Sn obtained from this theory are much better
than those obtained from the LSSM, while
those for even-even $^{116-128}$Sn obtained from the LSSM are a little better.
The corresponding ground-state $s$-boson occupation probability $\rho_{s}$
for these nuclei is shown in FIG. \ref{fig3}(b),
which indicates that $^{118,120}$Sn, according to this theory,
are within the critical region of the BCS-BEC crossover
with the total number of $s$-bosons and
neutron pairs a little larger than the half-filling point value.

\begin{figure}[htp]
\begin{center}
\includegraphics[scale=.2795]{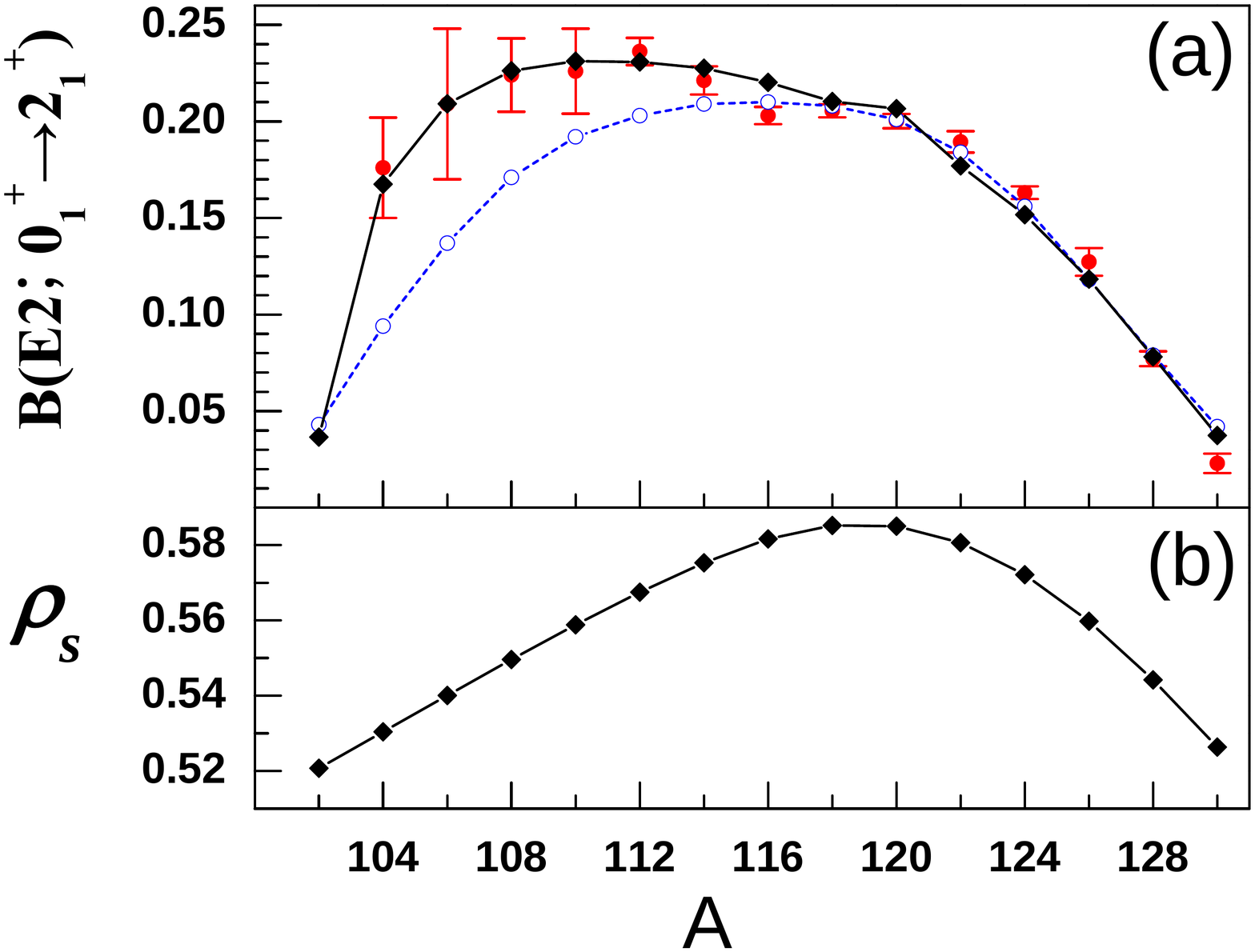}
\caption{(Color online) {\bf (a)} B(E2; $0^{+}_{1}\rightarrow 2^{+}_{1}$) values (in e$^{2}$b$^{2}$)
of even-even $^{102-130}$Sn obtained
in this theory
and in comparison with
the experimental data provided in \cite{th2} and the large-scale shell
model (LSSM) results \cite{be20},
where the solid dots with error-bar are the experimental values,
the solid diamonds linked by a solid line are the values obtained in this theory,
and the open circles linked by a dashed line are those calculated from the LSSM.
{\bf (b)} The ground-state $s$-boson occupation probability $\rho_{s}$ in even-even
$^{102-130}$Sn predicted in this model.
\label{fig3}}
\end{center}
\end{figure}

\vskip .5cm
\noindent{\bf Conclusion:}
In conclusion, the BCS-BEC coexistence and crossover phenomena
in the nuclear mean-field plus standard pairing interaction
model involving the $s$-bosons
is observed from the analysis of
the ground state $s$-boson occupation probability.
It is shown that, though the BCS-BEC crossover behavior may be different
from one Pauli blocking configuration to another,
the BCS-BEC coexistence seems robust, as demonstrated in the analysis
for the $pf$-shell with realistic single-particle energies
in two different blocking configurations or in the
nucleon-pair orbit and the $s$-boson independent approximation.
The crossover is further enhanced with increasing pairing interaction
strength.
In the example of the model application to even-even $^{102-130}$Sn,
the B(E2) values are well reproduced by the model
in the nucleon-pair orbit and the $s$-boson independent approximation
with parameters determined by the experimental $2^{+}_{1}$
level energy. A mixed state consideration related to the Pauli blocking
effect in the model
merits further study. Moreover, since BCS-BEC coexistence
seems common in nuclei, similar to the interacting Boson-Fermion model for
odd-A nuclei~\cite{ibfm}, in which, besides the $s$- and $d$-bosons,
only a single nucleon degree of freedom
is considered, a configuration which considers both the
$s$- and $d$-bosons provided from the IBM
and a few interacting valence nucleons
in an effective  mean-field, such as the shell model,
may be a better description of the low-energy structure of nuclei.
Further work along these directions is in progress.

\vskip .3cm
\begin{acknowledgments}
{Support from the National Natural Science Foundation of China (11375080 and 11675071),
the  U. S. National Science Foundation
(OCI-0904874 and {ACI-1516338}),
{U. S. Department
of Energy (DE-SC0005248)}, the Southeastern Universities Research Association,
the China-U. S. Theory Institute for Physics with Exotic Nuclei (CUSTIPEN) (DE-SC0009971),
and the LSU--LNNU joint research
program (9961) is acknowledged.}
\end{acknowledgments}

\clearpage

\end{document}